# "I had a solid theory before but it's falling apart": Polarizing Effects of Algorithmic Transparency

Aaron Springer[†]
Computer Science
University of California Santa Cruz
Santa Cruz, CA, USA
alspring@ucsc.edu

Steve Whittaker
Psychology
University of California, Santa Cruz
Santa Cruz, CA, USA
swhittak@ucsc.edu

## ABSTRACT

The rise of machine learning has brought closer scrutiny to intelligent systems, leading to calls for greater transparency and explainable algorithms. We explore the effects of transparency on user perceptions of a working intelligent system for emotion detection. In exploratory Study 1, we observed paradoxical effects of transparency which improves perceptions of system accuracy for some participants while *reducing* accuracy perceptions for others. In Study 2, we test this observation using mixed methods, showing that the apparent transparency paradox can be explained by a mismatch between participant expectations and system predictions. We qualitatively examine this process, indicating that transparency can undermine user confidence by causing users to fixate on flaws when they already have a model of system operation. In contrast transparency helps if users lack such a model. Finally, we revisit the notion of transparency and suggest design considerations for building safe and successful machine learning systems based on our insights.

## CCS CONCEPTS

• Human-centered computing~Human computer interaction (HCI)

## 1 Introduction

Machine learning algorithms power intelligent systems that pervade our everyday lives. These systems make decisions ranging from routes to work to recommendations about criminal parole [2,7]. As humans with limited time and energy, we increasingly delegate responsibility to these systems with little reflection or oversight. Nevertheless, intelligent systems face mounting concerns about how they make decisions; concerns that are exacerbated by recent machine learning advances like deep learning that are difficult to explain in human-comprehensible terms. Major public concerns have arisen following demonstrations of unfairness in algorithmic systems with regards to gender, race, and other characteristics [8,66,70]. The need for explanation and transparency is a core problem that is threatening adoption of intelligent systems in many realms [28,57,74].

There are many important reasons why algorithmic transparency is needed. Greater transparency can potentially increase user control and improve acceptance of complex algorithmic systems [38]. It can also promote user learning and insight from complex data, as humans increasingly work with complex inferential systems for analytic purposes [38,64]. Transparency can also enable important oversight by system designers. Without transparency it may be unclear whether an algorithm is optimizing the intended behavior [29,47], or whether an algorithm has negative, unintended consequences (e.g. filter bubbles in social media; [9,58]). Given these issues, it is increasingly possible that transparency, "a right to explanation", may become a legal requirement in some contexts [25]. These points have led some researchers to argue that machine learning must be 'interpretable by design' [1,40], and even essential for the adoption of intelligent systems, such as in cases of medical diagnoses [28,74].

While such calls for transparency are admirable, it is unclear exactly what is needed to enact them in practice. Extensive research about *how* to operationalize transparency has risen from both machine learning and HCI communities but no clear consensus has resulted [1,18,73]. The *how* of transparency is difficult—there are numerous implementation trade-offs involving accuracy and fidelity. Making a complex algorithm understandable to users might require explanatory simplification, which often comes at the cost of reduced accuracy of explanation [41,65]. For example, methods have been proposed to explain neural network algorithms in terms of more traditional machine learning approaches, but these explanations necessarily present approximations of the actual algorithms deployed [49].

In addition, recent empirical studies have also attempted to present a case for the *why* of transparency; however, these studies have shown puzzling and sometimes contradictory effects. In some settings there are expected benefits: transparency improves algorithmic perceptions because users may better understand system behavior [37,38,45]. But in other circumstances, transparency can have other quite paradoxical effects.

Transparency may cause user to have worse perceptions of a system, trusting it *less* because the transparency led them to question the system even when it was correct [45]. Providing system explanations may also undermine user perceptions when users lack the attentional capacity to process complex explanations for example while they are executing a demanding task [11,75]. There is mixed evidence for user *why* transparency should be implemented for users.

These results indicate that we lack a clear explanation of why transparency has seemingly contradictory effects. The current study aims to provide such an explanation. Rather than focusing on *how* to operationalize transparency or *why* to be transparent, we frame the problem differently: *when* is the best time to present transparency? We approach this question using empirical mixed-methods to understand transparency in the context of a working algorithm that interprets a user's description of an emotional experience. In two studies, we explore the connection between the impact of transparency and the user/system context it is presented in. We identify instances *when* transparency increases users' understanding and confidence, as well as *when* it might undermine user confidence. We examine the relationship between user expectations, system output, and transparency. We address the following research questions (RQs):

- (RQ1): When does transparency help versus hinder users' perceptions of complex systems? (Study 1 & 2)

- (RQ2): Why does transparency help or hinder user understanding? (Study 1 & 2)

- (RQ3): How does transparency influence users' expectations and perceptions of system error? (Study 2)

To answer these questions, we conducted two studies. Study 1 explores how users engage with and understand a transparent working intelligent system using 'think-aloud' and semi-structured interviewing methods. Study 2 takes our observations from Study 1 and incorporates a quantitative design to test predictions regarding how users form perceptions of accuracy in the intelligent system. We then explain these results in terms of expectation violation and theories of social explanation.

## 1.1 Contribution

We contribute to the growing literature on algorithmic transparency through two user evaluations of a working intelligent system in the Personal Informatics domain. Previous research on transparency and intelligibility has had highly mixed results. Positive system perceptions can be built through transparency [17,37,46], even to the point of overconfidence [23]. At the same time, however, positive system perceptions can also be undermined by transparency [20,37,45,54]. Our approach draws on psychological and sociological theories of communication applied to HCI [24,26,59,69] to explain when, why, and how users want transparency. Our findings reveal that transparency can have both positive and negative effects depending on context. We present a model that shows how the context of transparency and expectation violation interact in forming user perceptions of system accuracy. Transparency information has positive effects both in helping users form initial working models of system operation and reassuring those who feel the system is operating in unexpected ways. At the same time, negative effects can arise when transparency reveals algorithmic errors that can undermine confidence in those who already have a coherent view of system operation. We explain our results using theories of occasioned explanation [24,26], arguing that transparency information is anomalous for users who feel the system is operating correctly and therefore undermines their confidence in the system. Design implications include a greater focus on what situations necessitate a transparent explanation as well as improved algorithmic error presentation.

## 2 Related Work

## 2.1 Folk Theories of Algorithms

A wealth of prior work has explored issues surrounding algorithm transparency in the commercial deployments of systems for social media and news curation. Social media feeds are often curated by algorithms that may be invisible to users (e.g., Facebook. Twitter, LinkedIn). At one point, most users were unaware that Facebook newsfeeds were not simply all the posts that their friends made [21]. These users reacted in surprise and sometimes anger when they were shown the posts that were missing from their newsfeed. Later research shows that many users of Facebook develop 'folk theories' of their social feed [19], which are imprecise heuristics about how the system works, even going so far as to make concrete plans based upon their folk theories. This work also showed that making the design more transparent or seamful, allowed users to generate multiple folk theories and more readily compare and contrast between them [19].

Other work has illustrated issues regarding incorrect folk theories in the domain of intelligent personal informatics systems, showing specific challenges in how users understand these systems. Users are prone to blindly believing outputs from algorithmic systems, a phenomena referred to as algorithmic omniscience [20,32,67] and automation bias [15,53]. For example, KnowMe [72] is a program that infers personality traits from a user's posts on social media based on Big Five personality theory. KnowMe users were quick to defer to algorithmic judgment about their own personalities, stating that the algorithm is likely to have greater credibility than their own personal statements (e.g., "...*At the end of the day, that's who the system says I am...*"). Similar results were shown in [32], showing that participants expected intelligent personal informatics systems to serve as ground truth for their experiences and even attributed superhuman qualities to these devices, e.g., "...*[it] could tell me about an emotion I don't know that I am feeling...*". Other experiments indicate the risk of such trust, showing that users may believe even entirely random system outputs as moderately accurate [67]. Similarly, giving users placebo controls over an algorithmic interface shows similar results [71]; when given placebo controls, users felt more satisfied with their newsfeed.

Without a standard of transparency in intelligent systems, it may be easy to deceive end-users into believing they are using a real system; this is a dangerous proposition when apps can be so easily distributed.

## 2.2 Transparency and Expectation Violation

There is a long history of studying transparency and intelligibility in automated systems [5]. However, the results have mixed and often indicate contradictory effects. Many experiments have indicated that transparency improves user perceptions of the system [17,46]. Others have shown that interventions that simply show prediction confidence improve users system perceptions [3]. In extreme cases, animations that simulate apparent transparency can cause users to be overconfident about systems even when they err [23].

Other studies show less positive effects for user perceptions of a system. Participants who completed an experiment using a hypothetical transparent system were led to question the system an increased amount, resulting in worse agreement with the system [45]. However, the effect may be opposite for high certainty systems—transparency may only result in higher user agreement. Muir and Moray conclude that any hint of error in an automated system will decrease trust [54]. More recent work indicates other effects; explanations of how a system is working may lead to increased trust [37] but further explanations may be harmful to user perceptions. However, these effects are dependent upon the amount of *expectation violation* that a user experiences. User expectation violation follows an event where a system behaves in a way that a user did not expect [37,67,71]. To use an example from Kizilcec, if a student in an online course is confident of their abilities and the correctness of their work, they may experience expectation violation when they receive an automated test score that is very low.

Recently, the machine learning community has begun grappling with issues of transparency and explainability. This seems due to the rise of more inscrutable methods like deep learning as well as legal requirements arising from the European Union's GDPR . Some machine learning models are "inherently understandable" such as linear models and Generalized Additive Models [48,73]. These understandable models can be "explained" to users simply through the linear contributions of their features. Other algorithms such as deep neural nets and random forests are inscrutable, and it is nontrivial to explain how input features match to output predictions [73]. Many attempts have been made to make these inscrutable algorithms understandable. These rely on approximating the inscrutable algorithm through a simple local or linear model than can be explained to the end user [49,64]. However, it is not clear when users want such explanations or how these explanations should be framed. Many such AI attempts at transparency are not tested with users or have simulated user studies [4,52,55]. We must first learn when and how users will benefit from such transparency before we can begin to apply the techniques that have been developed.

## 2.3 Explanation and Persuasion Theory

People interact with computers and intelligent systems in ways that directly mirror how they interact with people [56,63]. Therefore, we turn to psychology and sociology for theory to inform how transparency should operate. Transparency allows for explanation of why a model made a given prediction. Therefore, we can turn to fields such as psychology and sociology for guidance about operationalizing explanations; these fields have a long history of studying explanation. Hilton shows that causal explanation takes the form of conversation and thus is governed by the common-sense rules of conversation [30]. Grice previously elucidated these rules, stating that participants communicate together following implicit rules known as the "Cooperative Principle" [26]: The cooperative principle consists of 4 maxims: Quantity, Quality, Relation, and Manner. Most relevant to us is Quantity—"1. Make your contribution as informative as required. 2. Do not make your contribution more informative than is required." Also of interest are Garfinkel's breaching experiments where social norms are violated [24]. One example of this is an experimenter asking for repeated explanation of commonplace phrases like responding to a "How are you?" from an acquaintance with "How am I with regard to what? My health, my finances, my school work, my peace of mind, my …?" These breaching experiments caused exasperation and social sanctions from those who experienced them. It seems likely that people will similarly sanction systems who breach social norms of explanation.

Additionally, we see parallels between how people interact with intelligent systems and persuasion. The Elaboration Likelihood Model (ELM) is a dual process model of persuasion [59]. The ELM posits that two parallel processes are engaged when a person evaluates an argument, similar to Kahneman's conception of System 1 and 2 thinking [34]. The central processing route involves careful consideration of the argument and complex integration into a person's beliefs. The central route is often engaged in high stakes decisions. The peripheral route in contrast focuses on heuristic cues such as the attractiveness of the speaker, the person's current affect, the number and length of the arguments, and other cues not directly related to the content of the argument. Prior work on intelligent systems seems to align with this dual process model [37], people understand systems through peripheral routes if their expectations are met, only engaging in central processing when their expectations are violated. This is also demonstrated this in the context of Google search suggestions; where users felt the cost of processing explanations outweighed their benefits [11].

## 2.4 Emotional Analytics

Our focus is on how users interact with an intelligent personal informatics systems which are being increasingly deployed within commercial [77–80] and research domains [6,22,31,50,76]. These systems track how a person operates on some dimension, whether physical, emotional, or mental, and then can suggest improvements to this behavior through customized feedback and suggestions [31,61]. Such data potentially allows users to analyze and modify their behaviors to promote well-being [12,13,33].

Furthermore, in contrast to other work that presents hypothetical scenarios in which participants read about or watch algorithmic deployments and decisions [23,45], our aim was to have users experience the algorithm *in situ,* as it directly made decisions about their own data [36]. One important characteristic of emotional interpretation is that users are knowledgeable about the status of their own feelings and experiences, allowing them to directly compare algorithmic interpretations with their own personal evaluations of their emotional experiences. This contrasts with other applications of smart algorithms, such as medical diagnoses. In these complex realms regular users might be less able to interpret the results of algorithmic interpretations. In addition, emotion is highly variable between individuals and previous research demonstrates difficulty in accurately predicting emotion from text [43,62]. This allows us to closely examine how users understand a system under varying degrees of error.

## 3  Research System: E-meter

We developed a working system called the E-meter that uses textual entries to predict emotion. The E-meter (Figs 1,2) presents users with a web page showing a system depiction, a short description of the system, instructions, and a text box to write in. The system was described as an "algorithm that assesses the positivity/negativity of [their] writing".

The algorithm underlying emotion detection worked in the following way: each word that was written by the user was checked for its positive/negative emotion association in our model. If it was found in the model, the overall mood rating in the system was updated. This constitutes an incremental linear regression that recalculates each time a word is written.

### 3.1  Machine Learning Model

As we outlined in the background, current processes for explanation of inscrutable models such as deep neural networks involve approximating the inscrutable model by a simpler, often linear, model [73]. Therefore, we focus on a linear model so that our transparency can be operationalized in a way that is faithful to current research.

Emotion predictions for users' experiences were generated using a linear regression model trained on text from the EmotiCal project [31,68]. In EmotiCal, users wrote short textual entries about daily experiences and evaluated their mood in relation to those experiences. This data gave us a gold-standard supervised training set on which to train our linear regression. We trained the linear regression on 6249 textual entries and mood scores from 164 EmotiCal users. Text features were stemmed using the Porter stemming algorithm [60] and then the top 600 unigrams were selected by F-score, i.e. we selected the 600 words that were most strongly predictive of user emotion ratings. Using a train/test split of 85/15 the linear regression tested at $R^2 = 0.25$; mean absolute error was .95 on the target variable (mood) scale of (-3,3). In order to implement this model on a larger range for the E-meter, we scaled the predictions to (0,100) to create a more continuous and variable experience for users. The mean absolute error of our model indicates that the E-meter will, on average, err by 15.83 points on a (0,100) scale for each user's mood prediction.

Version 1: **Document-level:**

As users wrote, the E-meter showed the system's interpretation of the emotion of their writing. If the overall text was interpreted as positive, the meter filled the gauge to the right and turned more green (Fig 2); if the text was interpreted negatively, the gauge was emptied to the left and turned more red (Fig 1). This feedback represents the coarse and global feedback that many machine learning systems currently display. These systems give an overall rating but don't allow the user insight into the detailed workings of the algorithm.

Version 2: **Word-level**

In contrast with the Document-level version, the word-level condition provided fine-grained transparency. We operationalized transparency by highlighting the mood association of each word in the model; if a word is highly associated with a positive mood then it will be highlighted green, a word associated with a negative mood will be highlighted orange or red. The word-level version showed immediate incremental feedback of how the system interpreted each word as the user types. In this word-level condition, individual words are highlighted and color coded according to how the underlying algorithm interpreted that word's affect. This incremental feedback allows users to see how each individual word they wrote contributed to the overall E-meter rating. Furthermore, words remained highlighted as users continued to type allowing them to continue to assess their contribution to the overall score.

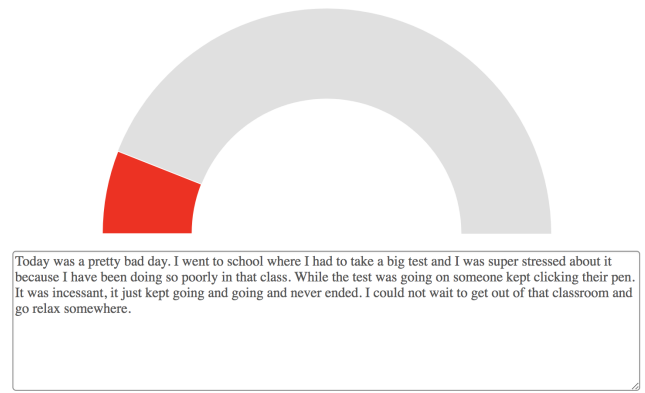

**Figure 1: E-meter Document-Level Feedback Condition**

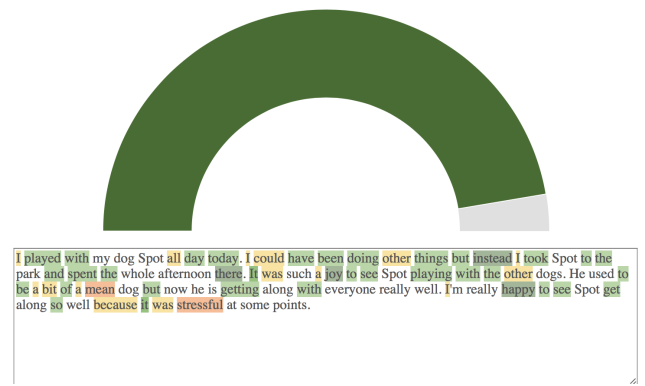

**Figure 2: E-meter Word-Level Feedback Condition**

This form of transparency offers users insight into the underlying word-based regression model driving the E-meter visualization; it depicts how the regression model correlates each word with positive or negative emotion to arrive at an overall weighting for the entire text that the user has entered. The fact that the visualization is persistent also allows users to reexamine what they have written, reconciling the overall E-meter rating with the fine-grained word-level connotations.

We could have operationalized transparency in other ways. Other researchers have operationalized transparency through natural language explanations [37] and diagrams [45]. However, in our case we can convey nearly the entire working of the system through word highlighting. In addition, our operationalization allow the answering of counterfactual questions, an important part of explanation [51,73]. Highlighting the text is a non-intrusive way of conveying to the user what drives the algorithm and gives direct clues about the underlying linear model. In addition, by varying the colors of the highlighting we also show how the model is interpreting the specific words.

## 4 Study 1

Our first exploratory study aimed to understand the processes by which participants make sense of a complex algorithm that interprets their emotions.

### 4.1 Method

*4.1.2 Users.* Twelve users were recruited from an internal participant pool at a large United States west-coast university. They received course credit for participation. Participants average age was 19.54 years (sd=1.52) and 7/12 identified as female. This study was approved by an Institutional Review Board.

*4.1.3 Measures.* All survey questions requested Likert scale responses unless stated otherwise. Participants were asked about their experience with the E-meter including: "Select the number of times you looked at the visualization while you were writing" and subsequently "If you looked at the visualization more than once: Rate the extent to which looking at the visualization impacted or did not impact your writing." We next probed user evaluations of system accuracy and their trust in the system: "How accurate or inaccurate did you find the E-meter?" and "How trustworthy or untrustworthy did you find the E-meter system?" In addition to these questions, we used a shortened version of the Psychological General Well-Being Index (PGWBI) to screen for mental health before participants began the study [27].

Additionally, we assessed users' perceptions of the emotion of their writing: "How positive or negative did you feel our writing was?", as well as the system's evaluations of their writing: "How positive or negative did the E-meter assess your writing to be?" We used the absolute difference of these two measures to calculate an aggregate measure of *expectation violation*. If the E-meter were perfect, it would always predict exactly how the user felt and expectation violation would be 0. If a user felt that their writing was "Strongly Negative" (1) but the E-meter rated it as "Slightly Negative" (3) then the user's expectation violation would be 2.

*4.1.4 Procedure.* The participants were randomly divided into one of two conditions. Both groups were given document-level affective feedback from the E-meter scale as shown in Fig 1.

- Condition 1: Six participants received real-time incremental word-level feedback about the algorithm's interpretation of their affect as they typed each word.

- Condition 2: The other six only obtained word-level feedback after they had finished the writing task; these users explicitly requested word-level feedback by clicking a button labeled "How was this rating calculated?".

The researcher explained the experiment and think-aloud procedure, demonstrating a think-aloud on an email client. The researcher asked participants to "Please write at least 100 words about an emotional experience that affected you in the last week." In cases where the participant had trouble thinking aloud, they were prompted to speak. After the think-aloud writing exercise, the experimenter conducted a semi-structured interview that included an on-screen survey that the participants continued thinking aloud as they answered. After the survey, participants in the word-level feedback condition 2 were presented with the final state of the E-meter, exactly as they saw it when they finished writing. Participants in the initial document-level condition 1 saw exactly the same screen with an added button labeled "How was this rating calculated" which they pressed to reveal word-level highlighting. Finally, all participants were presented with a printed version of the final E-meter state with which they marked up to indicate errors. The entire process took around 50 minutes.

*4.1.5 Analysis.* Interviews were recorded using both audio and screen-recording. Two interviews (one from each condition) were not audio recorded, thus only the remaining 10 are used for the analysis. Recall RQ1 for this study: When does transparency help versus hinder users' perceptions of complex systems? We analyzed the interviews with RQ1 in mind; responses were coded using theoretical thematic analysis [10]. We describe the major themes related to RQ1 below.

### 4.2 Study 1 Results

Participants engaged meaningfully with the feedback from the E-meter—all participants consulted feedback at least once and 8/12 of participants consulted it 'more than 5 times'.

*Document-Level Feedback Results in Inaccurate Mental Models:* Only half of the participants receiving document-level feedback formed accurate mental models of the system. Others with document-level feedback expressed confusion about how the algorithm operated, which in turn negatively affected their system perceptions. Participant 10 expressed this confusion and how it led them to consider the system as inaccurate:

*P10: I don't have a way of interpreting it, I wouldn't know if it's good or bad or what I'm writing is negative or positive…. it's inaccurate cause it doesn't seem to portray the true emotional state of what I wrote, and untrustworthy cause it doesn't give off the right feedback, it doesn't allow me to interpret it correctly.*

Participant 11 who also received document-level feedback considered the E-meter may be working entirely off of *"tone in my voice"* or at random *"…this could have been a whole fake fluctuation."*

*Word-level feedback Promotes More Accurate Mental Models:* When we later showed word-level feedback to these document-level participants, their confusion seemed to dissipate, leading them to form more accurate mental models about how the algorithm worked. In this quote from Participant 10, note the stark difference from their prior quote above; they now are reassured that the system is actually working and make excuses why the system generated an incorrect rating.

*P10: it goes word by word, it tries to take positivity and negativity from each word, I don't think it really goes by context of what I'm writing much more than the word itself, so saying something like 'my best friend getting arrested', it's definitely in a negative light, but because I mention 'my best friend', it kinda took it in a positive light.*

In contrast, many participants who received word-level feedback throughout formed accurate mental models initially when using the system. Participant 5 said:

*P5: I think it tags certain words, for sure, with values probably 1-4, green, yellow, orange, red, or nothing, that there are certain words that are programmed into it. … Maybe there's words that it knows to look for across the system, like 'death,' 'burial,' negative words.. 'celebration,' 'party,' it can pick up on across the whole thing.*

Word-level transparency information therefore seemed to help both groups by promoting insight into how the algorithm was making its evaluations, either in real-time or retrospectively. These observations suggest overall benefits for word-level information, confirming our initial expectations about the value of providing this type of transparency feedback. Despite these benefits, to our surprise, we also observed some *negative* effects for word-level feedback, which on some occasions seemed to undermine some participants' views of the algorithm.

Participant 3 when examining the word-level feedback after they had written their text noted *"…honestly I feel like I don't see a pattern at all and it's kind of bugging me."*

Participant 8 first used the document-level feedback system and formed a mental model of system operation. Upon being shown the word-level feedback they began to question their established prior model; this resulted in worse perceptions of the system:

*P8: "Yeah, I'm actually not sure what—I don't know if it's just as simple as positive versus negative words. **I had a solid theory before, but it's falling apart...** Well I just assumed that some words would be coded as positive or negative and then it would just like do a ratio of those two."*

Overall, it seems that word-level transparency offered people a useful heuristic, but this heuristic could be undermined by closer analysis.

*Expectation Violation Predicts Accuracy Perceptions:* We also wanted to quantitatively check whether participants evaluations of accuracy related to their expectation violation. As expected, we see a strong correlation between the amount of expectation violation that users experienced in the visualization and their overall perception of the system's accuracy ($r(11)=-0.748$, $p=0.005$). In other words, users judge the system as accurate if the system's interpretation is consistent with the user's evaluation of their writing.

### 4.3 Study 1 Summary

Transparency feedback seemed to provide a useful heuristic in helping users form working mental models, but closer scrutiny and perceived errors may undermine confidence in the system. When users felt the system was inaccurate in the document-level condition, they were reassured later by word-level feedback showing how the system actually worked. However, we also see positive system perceptions undermined by the word-level feedback when users had established mental models. We set out to explore these seemingly contradictory effects in a larger scale quantitative study. Again, we compared both forms of feedback, but we also wanted to more directly explore potentially differing effects of word-level feedback in relation to expectation violation. How might word-level feedback both (a) help users who were unclear about how the algorithm operated while at the same time (b) reduce the confidence of users who already had a working theory that was subsequently undermined when they were confronted with word-level errors?

## 5 Study 2

### 5.1 Method

We used the same system and conditions as Study 1 with one important difference. While users in the Document-Level feedback condition in Study 1 eventually saw the transparent Word-Level feedback after they had completed their writing, in Study 2 the conditions are entirely separate, Document-Level users never see Word-Level feedback. Users were randomly divided into Word-Level and Document-Level conditions and instructed to write 100 words about an emotional experience in the E-meter system.

*5.1.1 Users.* We recruited 41 users to test the E-meter system who had previously passed a short mental health screening (PGWBI) [27]. Users were recruited from Amazon Turk and paid $3.33. The evaluation took 13 minutes on average. This study was approved by an Institutional Review Board.

*5.1.2. Measures.* We asked the same questions as Study 1 with the addition of the following questions. "Please name 2 or more things you like about the system" and "Please name 2 or more things you dislike about the system". "Please give 2-3 ways the algorithm affected your writing", "Imagine that you were given personalized tips on how to improve your mood based on what you wrote. Would you make use of such suggestions?", "Please explain how do you think the system judges your writing.", ""Did you experiment with or manipulate your writing to test how the system was working or how accurate it was? If so, how?", "If you have any additional feedback from your interaction with the E-meter, please detail it here."

## 5.2 Study 2 Results

Participants followed the instructions to write at least 100 words, mean=107.74, sd=14.8. As shown in Fig 3, the majority of users across conditions found the E-meter to be "Accurate" or "Very Accurate" with the median being "Accurate". Fig 4 shows that users found the E-meter to be "Moderately Trustworthy".

As in Study 1, we calculate a user's expectation violation in relation to the system's overall emotion rating. We find that transparency and expectation violation interact in a complex manner. We see a strong negative correlation between expectation violation and accuracy in the document-level group, (r(21) =-.898, p < .00001) confirming Study 1, as well as prior work [67]. In other words, with document-level feedback, when the system behaves as the user expects then it is perceived as accurate. However, this correlation between expectation violation and accuracy perceptions disappears in the word-level condition: r(16) = -0.175, p = 0.488. The relationship between expectation violation and accuracy perceptions is clearly more complex in the presence of word-level transparency.

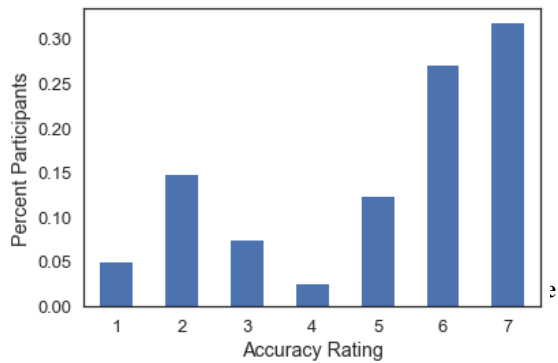

**Figure 3: Participants found the E-meter Accurate**

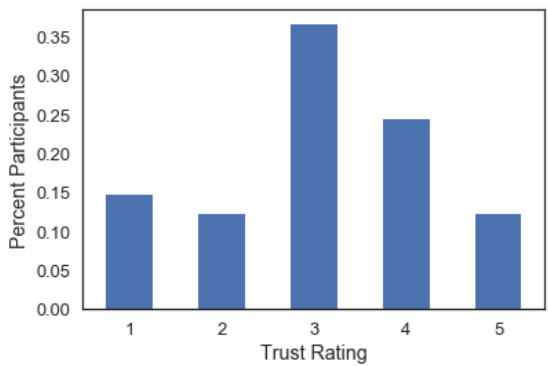

**Figure 4: Participants found the E-meter Moderately Trustworthy**

|  | Coef | SE | p-value |
|---|---|---|---|
| Intercept | 6.991 | 0.377 | **< 0.0001** |
| Expectation Violation | -1.736 | 0.601 | **< 0.0001** |
| Condition | -1.406 | 0.198 | **0.007** |
| Condition * Expectation Violation | 1.057 | 0.441 | **0.022** |

**Table 1: Effects of Transparency on Perceived Accuracy. ($R^2$ = .55, p < .0001).**

We modeled the effects of the different types of transparency more systematically using linear regression predicting user accuracy perceptions as dependent variable (see Table 1). Given that Study 1 indicated people may respond to transparency differently based on expectation violation, independent measures in the regression model include expectation violation, condition, and an interaction between them. The overall regression was highly predictive $R^2$=.548, p < .0001. As expected, both transparency and expectation violation are associated with perceived accuracy. Counterintuitively, adding word-level transparency has an overall *negative* effect on perceptions of accuracy. However, this overall effect depends on expectation violation, as indicated by the interaction term in the regression. We depict the interaction in Fig 5, which shows that when compared with document-level feedback, word-level transparency has the expected positive effect when expectation violation is high. Confirming our qualitative results from Study 1, people using the word-level version of the system show higher levels of perceived accuracy when their expectations aren't met. However, the effects of word-level transparency are negative for lower levels of expectation violation, when compared with document-level feedback. Thus, word-level transparency is unhelpful when people perceive the system to be accurate.

One possible explanation could be that users in the Word-Level condition were consciously modifying and going back to edit their writing in order to achieve an accurate overall rating. Recall that we asked users whether they had modified their writing according to the feedback from the E-meter. Seventeen of our 41 users said that they changed their writing in order to influence the E-meter's response. However, this did not seem to modify perceptions of accuracy of the E-meter. Adding a binary variable for modifying writing based on the feedback into our above regression (Table 1) was insignificant b=-0.23, p=.61. It seems that users' accuracy perceptions are not meaningfully affected by testing the E-meter.

To understand these effects further, we explored participant's qualitative responses to better understand the interaction between transparency and expectation violate on.

*5.2.1. Document-Level Feedback Results in Vague Mental Models*: Most users in the document-level feedback condition were focused on major shifts in the movement of the meter in relation to their writing. For some of these users, document-level feedback was consistent with their expectations, helping them to form a global, if vague, model of how the algorithm was operating. For example, P24 wrote about how the accuracy of the system related to the system meeting their expectations: *"I was writing about a negative topic and it continued to read in the negative state. The more upset I was writing, the further the dial went into the red."* Users for whom the document level feedback matched their expectations maintained high confidence that the system was accurate.

In contrast, other document-level users drastically lowered their confidence in the system's accuracy when the felt that the system outputs contradicted their expectations. P4 felt that the system was inaccurate overall but could not form a clear hypothesis about exactly why it was failing: *"It was highly inaccurate because the experience was clearly a negative one, I specifically explained how awful I felt, I don't think that it could measure the sentiment of what I'm writing."* Others also thought that the system was inaccurate, but the absence of transparency led them to speculate about other ways it could be working. Participant 11 said *"it looked at length and speed of what I was typing";* Participant 19 concurred, saying *"i thought it was only reacting to my WPM [words per minute]".*

Overall, document-level users are confident in system operation if their expectations are met. If their expectations are violated, they seem to doubt the system is working as indicated and, lacking any reassurance, this undermined their confidence in the system.

*5.2.2. Word-Level feedback effects depend on expectations*: Users in the word-level feedback condition formed clear mental models of how the system was working in rating individual words to arrive at an interpretation. However, even though these users seemed relatively confident about the algorithm's operation, they were often distracted or undermined by the system's interpretations of particular words. Participant 28 explained their rating of the system's accuracy as being downgraded by specific errors: *"It went way down when I typed the word "mad" but that was only a small part of the whole situation. The words that were good or bad seemed kind of arbitrary too."* Similarly, participant 30 said: *"I wrote, 'I was not thrilled' which is a negative statement, but this meter took the word 'thrilled' as a very positive thing."*

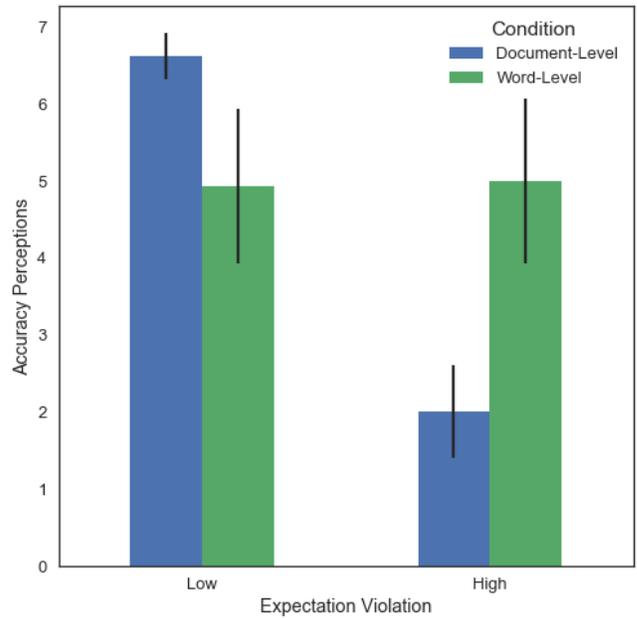

**Figure 5: Expectation Violation and Transparency Condition Interact to Form Accuracy Perceptions**

These examples suggest that, paradoxically, word-level errors might undermine some participants whose expectations are met and already have a good working theory of the algorithm's operation. To assess this further we examined cases of low expectation violation (instances of where users rated the E-meter as within 1 point of their own evaluation of their writing). Overall for these users the system is operating as expected. We analyzed these low expectation violation users to see why presenting word-level transparency information *reduced* perceptions of system accuracy.

For these users, word-level transparency seems to create more questions than it answered; the additional information provided by the word-level highlighting seemed to confuse rather than clarify. One participant noted that while the system's final negative rating was consistent with their overall judgement of their own writing, the highlighting didn't make sense *"...because the rating did not correspond to the number of identified words";* this user also noted *"It gave a positivity rating of 1 even though it only highlighted one or two words as red."* The word-level highlighting revealed to other users that the model worked in a different way from the user themselves. While users in the document-level condition were not able to discern this as a problem, word-level feedback users took issue with this. Participant 41 said: *"The key is to measure the overall emotional tone of the passage and it seems to fail at this."* Participant 36 said this simply: *"I disliked that it cannot understand context."* For these users, transparency revealed that the algorithm did not conform to their mental models of the task.

We also examined users who had the opposite experience. We looked at users who initially felt that the algorithm was violating their expectations, but transparency seemed to help them. For them word-level transparency seemed to provide reassurance and explanation of the system behavior. One user, participant 27 seemed to note that the system was trying, even if it violated their expectations: *"I think that it was measuring words i used and rated them almost correctly."* In the same vein, Participant 30 said *"Even though it got several individual things wrong, I think it actually did a good job on the whole."*

Overall, word-level feedback seemed to have somewhat contradictory effects depending on the user's assessment of system performance. For users who felt that the system was behaving appropriately, noticing word-level errors and non-conformance to their mental models undermined their views of system operation. For others who were less sure about overall system accuracy, word-level feedback had the opposite effect, as it boosted confidence in the system. These data are consistent with Fig 5 showing that compared with document-level feedback, word-level feedback reduces accuracy judgements in low expectation violation users, but increases it for those who have high violations.

### 5.3 Study 2 Discussion

In this study we presented both qualitative and quantitative data showing that algorithmic transparency has complex effects that depend on users' expectation violation. Word-level transparency users with the most violated expectations had better perceptions of the E-meter's accuracy compared to their document-level counterparts. However, users in the word-level condition were less likely to regard the system as highly accurate when it did not violate their expectations.

## 6 Discussion

Given their widespread deployment, it is imperative that we derive new theories and design approaches to improve users' understanding of intelligent systems. Two studies show that fine-grained algorithmic transparency in a working intelligent system can lead to very different effects for different users. Study 1 showed that some users are reassured when they view word-level transparency; it shows them that the system is working even if it didn't generate exactly the interpretation they expected. However, for other users, transparency seemed to undermine their experience; while overall the system worked as they expected, they also detected specific word-level feedback that seemed anomalous. Study 2 resolves these apparently contradictory observations. We show that there is no one-size fits all solution, as responses to word-level transparency depend on users' level of expectation violation. In general, users whose expectations are violated strongly will be reassured by the word-level transparency. On the other hand, users who initially believe the system is operating accurately may begin to question their ratings if they are shown word-level transparency. This undermining effect seems to be strongly influenced by a user focus on system errors; despite an overall view that the system is operating as expected, seeing individual errors reduces confidence and system trust.

### 6.1 Limitations

The current study explores one algorithmic domain, emotion regulation and clearly other contexts need to be explored. Furthermore, our deployment of a working algorithm meant that results were obtained for situations where our algorithm generated moderate numbers of errors, and future research should evaluate contexts where there are different levels of errors, including extremes of multiple versus few errors. Additionally, while users generated their own data while using a real working system, results were not directly used to inform other aspects of user's personal behavior such as emotion tracking or emotion regulation, so the costs of system errors were low. While this is appropriate for exploring understanding of initial algorithms with moderate error rates, future work might explore user system models and trust in more high stakes contexts. Also, our work explores one aspect of transparency namely a dynamic visualization of the algorithm, and there are many other ways to depict how an algorithm operates including verbal explanations, concrete user exploration and so forth [37,39,44,64].

### 6.2 Synthesizing Contradictory Results

Our results also serve to synthesize and explain previous research on transparency and intelligibility that has generated highly mixed and sometimes paradoxical results. On the one hand, trust can be built through transparency [17,37,46], even to the point of overconfidence [23]. At the same time, however, trust can also be undermined by transparency [37,45,54]. Our findings indicate that this prior work may be rationalized by attending to expectation violation. For users who have low confidence in a system, with little idea how it works, then transparency can help form working system models, thus boosting confidence and trust. In contrast, trust can be undermined if users have a working theory of the system but are exposed to anomalous behaviors such as system errors.

### 6.3 Social Communication Theories to Support Transparency

The interactions between transparency and expectations are also consistent with social analyses of when explanations are needed. Theories of human communication (e.g. Garfinkel, Grice) argue that explanations are *occasioned*, i.e. that explanations are only provided on an 'as needed' basis when a situational expectation is not met [24,26]. According to these theories, transparency is therefore occasioned for expectation violations, making it contextually appropriate to provide an account for why system behavior is unexpected or unusual. But if expectations are met, i.e. the system and situation are perceived as going according to plan, then an explanation is anomalous and contextually marked, potentially reducing user confidence: 'everything is going fine so why are you providing this unnecessary information?'. Results are also consistent with the elaboration likelihood model [59]; which suggests that users might be generally happy to operate with imprecise working heuristics about how an algorithm operates, only invoking complex analysis when the algorithm behaves anomalously.

Our results draw attention to an important and often overlooked aspect of transparency. Most research has focused on questions of *how* to explain algorithmic operation, e.g. using approximate methods to explain neural nets using methods that everyday users will comprehend. In contrast our focus here has been on better understanding of *when* to deploy transparency, as we show that providing detailed information about an algorithm's operation can be counterproductive for users who have a working theory of system operation. Other work suggests that users do not want to be exposed to detailed algorithmic explanations when they are cognitively overloaded [32], suggesting a need to develop more systematic accounts of when algorithmic explanations are occasioned and useful.

Of course, deciding when to provide algorithmic explanations also gives rise to ethical concerns: if users are operating under false working assumptions about an algorithm then we need to expose and counteract these. Again, this suggests the need for an increased research focus on understanding users' working models of systems allowing us to diagnose when these are accurate, and when we need to intervene. This area is extraordinarily complex however given applications where positive placebo effects have been obtained using algorithms that falsely inform users that their stress levels are low [14]. In this case, an inaccurate user model and imperfect algorithmic understanding has beneficial outcomes.

## 6.4 Design Implications for Intelligent Systems

Our results also have implications for design. Whether or not transparency is beneficial depends on various characteristics of the application. For example, if an intelligent system is highly accurate overall in its predictions, then increasing the transparency may have a net negative effect of reducing perceptions of accuracy, given that users are highly attentive to even low levels of system error. If an application is inaccurate, then transparency could have a net positive by tempering those negative accuracy perceptions. Of course, other factors influence whether transparency is beneficial, such as the impact of the decision that the algorithm is influencing. When costs of errors are severe then it is imperative that user attention is drawn to these [16,42,81].

We find that in some contexts, transparency *decreases* perceptions of accuracy. One reason for this decrease seemed to be when transparency revealed that the way the system was operating was different from a user's mental model of the task. For example, users took issue with the fact that the system didn't seem consider context surrounding some words. Considering context was impossible in our machine learning model given that we were using solely individual words as features. Decisions made when training a machine learning algorithm including feature selection and algorithm choice necessarily constrain how transparency can be operationalized. This demonstrates that algorithmic decisions have direct user experience impacts in transparent applications. Current research indicates that we should present transparency in ways that bridge the gap between user's mental models and expert mental models of the task [18]. Our work dovetails with this, demonstrating that correct choices need to be made concerning the machine learning models to even allow for transparency that bridges this gap.

## 2.3 Presenting Errors in Intelligent Systems

Our results also draw attention to the critical problem of error presentation. All machine learning systems generate errors and while users need to be aware of this, our data suggests that in some cases users are overly focused on errors even when these are relatively infrequent and the system is operating well overall. This may be consistent with psychological theories suggest that people are generally poor at evaluating probabilities [35]. In any case, it indicates the need for much more research on error visualization as well as algorithmic success. Other designs suggested by our work could address the effects of errors undermining confidence. For example, we might only show highlighting on words that the system is very confident will be positive or negative in any context. Other users wanted to know about the relative weighting of positive versus negative words and the algorithm might provide more explicit models of this. These design improvements may improve perceptions of accuracy across the board allowing users to generate more stable models and reduced questioning. However, given the ubiquity of errors in all intelligent systems, much more research is needed to explore how errors might be presented and explained in ways that do not undermine the development of accurate working models of system operation.

For consumer facing applications it may be beneficial to operationalize transparency so as to promote user confidence in approximate system models allowing underlying complexity to be hidden from the user. In other contexts, however, we may want to have users very carefully evaluate a system prediction. For example, we may not want doctors to simply defer to the predictions of a medical decision support system. We may instead want to lead them to question their underlying model while using the system. For such cases, it may be beneficial to operationalize transparency to promote careful consideration with moderate amounts of expectation violation. This may be a difficult balance to achieve between questioning and validating a system model so that while skeptical users would continue to trust and use the algorithm.

Overall our results offer a new framework to synthesize paradoxical results surrounding transparency, explaining these in terms of social science theory. They also suggest new design implications when tackling this critical emerging area.

## ACKNOWLEDGMENTS
[anonymized]